\documentclass[%
twocolumn,
showpacs,
nofootinbib,
nobibnotes,
amsmath,amssymb,
aps,
pra,
longbibliography,
floatfix,
]{revtex4}
\usepackage{amsmath}
\usepackage{amsxtra}
\usepackage{amstext}
\usepackage{amssymb}
\usepackage{latexsym}
\usepackage{dsfont}
\usepackage{subfigure}
\usepackage{subfloat}
\usepackage{graphicx} 
\usepackage{epstopdf}
\usepackage{float}
\usepackage{dcolumn} 
\usepackage{bm} 
\usepackage{hyperref} 
\usepackage{natbib}
\usepackage{color}

\renewcommand{\vec}[1]{{\bf{#1}}}

\begin{document}\emph{}
\preprint{APS/123-QED}
\title
{$\mathbb{Z}_{2}$ topological insulator
 of ultra cold atoms in\\
  bichromatic optical lattices}

\author{Ahad K.  Ardabili}
\affiliation{Department of Physics, Ko\c{c} University, Sar{\i}yer, Istanbul, 34450, Turkey}

\author{Tekin Dereli }
\affiliation{Department of Physics, Ko\c{c} University, Sar{\i}yer, Istanbul, 34450, Turkey}

\author{\"{O} E.  M\"{u}stecapl{\i}o\v{g}lu}
\affiliation{Department of Physics, Ko\c{c} University, Sar{\i}yer, Istanbul, 34450, Turkey}

\begin{abstract}
We investigate the effect of a strong bichromatic deformation to the $\mathbb{Z}_{2}$ topological insulator in a fermionic ultracold atomic system proposed by B. B\'eri and N. R. Cooper, Phys.~Rev.~Lett. {\bf 107}, 145301 (2011). Large insulating gap of this system allows for examination of strong perturbations. We consider bichromatic perturbation along all axes on a triangular optical lattice. We find that $\mathbb{Z}_{2}$ topological character of the system is robust  up to a certain depth of the deformation. The lowest band can become topologically trivial while the lowest two bands are always protected.
\end{abstract}

\maketitle
\section{Introduction} \label{sec:introduction}
Topological insulators are
insulating in the bulk but have metallic states on their boundaries \cite{hasan,qi}.
Robustness of these states against disorder and perturbations makes them promising  for applications such as spintronics \cite{moore2010} and topological quantum computation \cite{nayak}. Topological invariants of the bulk material are essential for the robust boundary modes. This urged consideration of topological insulators on different lattice geometries \cite{lattices1,lattices2,lattices3,lattices4,lattices5}. Testing these systems against strong disorder and perturbations is challenging both experimentally and theoretically. Detailed numerical investigations reveal that topological protection is safe only for surface disorder smaller than the bulk band gap \cite{schubert}.

Two and three dimensional topological insulators with band gaps in the order of the recoil energy have recently been proposed in ultracold fermionic atomic gases \cite{z2cooper}.  This large gap systems are described in the nearly free electron limit, which is a different perspective to examine topological insulators compared to earlier studies \cite{imprint1,imprint2}. The proposal utilizes interactions which preserves time reversal symmetry (TRS), analogous to synthesized spin-orbit coupling  \cite{spielman},
so that the insulators are classified by the so called $\mathds{Z}_{2}$  topological invariant \cite{lattices5}.

Our aim in this article is to examine robustness of two dimensional $\mathds{Z}_{2}$ topological insulators \cite{z2cooper} against large global perturbations that breaks inversion symmetry but preserves TRS. Motivated by their recent use in
quantum simulations of relativistic field theories \cite{withaut,mazza2012} and disorder induced localization
\cite{aubry,roati,deissler2010}, we specifically consider here
bichromatic optical lattice perturbations. Similar studies of $\mathds{Z}_{2}$ topological insulators in
bichromatic potentials, but for a one-dimensional optical lattice in tight-binding limit, reveals
that topological properties of the system can be probed by density measurements \cite{mei2012,lang2012}.

This report is organized as follows. In Sec.~\ref{sec:model} we briefly
 review the proposal of $\mathds{Z}_{2}$ topological insulator in ultracold atomic gases \cite{z2cooper},
and introduce the bichromatic deformation of the lattice potential terms. In Sec.~\ref{sec:results} we describe the method of calculation of the
$\mathds{Z}_{2}$ invariant  \cite{kane} and present the
corresponding results. We conclude in Sec.~\ref{sec:conclusion}.
\section{Model system} \label{sec:model}
The Hamiltonian of an atom with $N$ internal states with position $\mathbf{r}$
and momentum $\mathbf{p}$ can be written as,
\begin{equation}\label{hamiltonian}
H=\frac{\mathbf{p}^{2}}{2m} +\hat{V}(\mathbf{r}),
\end{equation}
here $\hat{V}(\mathbf{r})$ is a position dependent potential which is a
$N \times N$ matrix acting on the internal states of the atom. A $\mathbb{Z}_{2}$
topological insulator is invariant under the action of time-reversal
 operator $\Theta= i \sigma_{y}\hat{K}$, where $\mathbf{\sigma}_{x,y,z}$ are Pauli matrices acting
 on electronic spin and $\hat{K}$ is complex conjugation operator.
 This requires $N$ to be even. The smallest  potential matrix has $N=4$ and can be written as,
  \begin{equation}
  \hat{V}(\mathbf{r})=\left(
    \begin{array}{cc}
      (A+B)\mathds{1}_{2} & C\mathds{1}_{2}-i \mathbf{\sigma}\cdot\mathbf{D}  \\
      C\mathds{1}_{2}+i \mathbf{\sigma}\cdot\mathbf{D} & (A-B)\mathds{1}_{2}  \\
    \end{array}
  \right).
  \end{equation}
Here $A$,$B$ and $C$ are real numbers, $\mathbf{D}$ is a 3-vector with real components
 and $\mathds{1}_{2}$ is $2\times 2$ identity matrix.

This Hamiltonian can be realized using an atom with four internal
states as proposed by \cite{z2cooper}. Ytterbium  $^{171}$Yb, which has
nuclear spin $I=2$, is a good candidate for this purpose.
It has a $2-$fold degenerate ground state $( ^{1}S_{0}=g)$ and
a long-lived $2-$fold degenerate excited state $(^{3}P_{0}=e)$;
furthermore there exists  a state dependent
 scalar potential for a specific wavelength
$\lambda_{\text{magic}}$ \cite{dalibard} in which the potential changes sign and becomes $\pm V_{\text{am}}(\vec{r})$
for the ground  and excited states.

\begin{figure}[t]
\subfigure[ ]{
\includegraphics[width=0.3\textwidth]{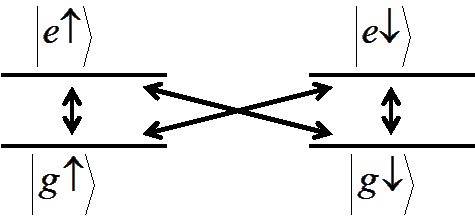}
\label{fig:fig1a}}
\qquad
\subfigure[ ]{
\includegraphics[width=0.4\textwidth]{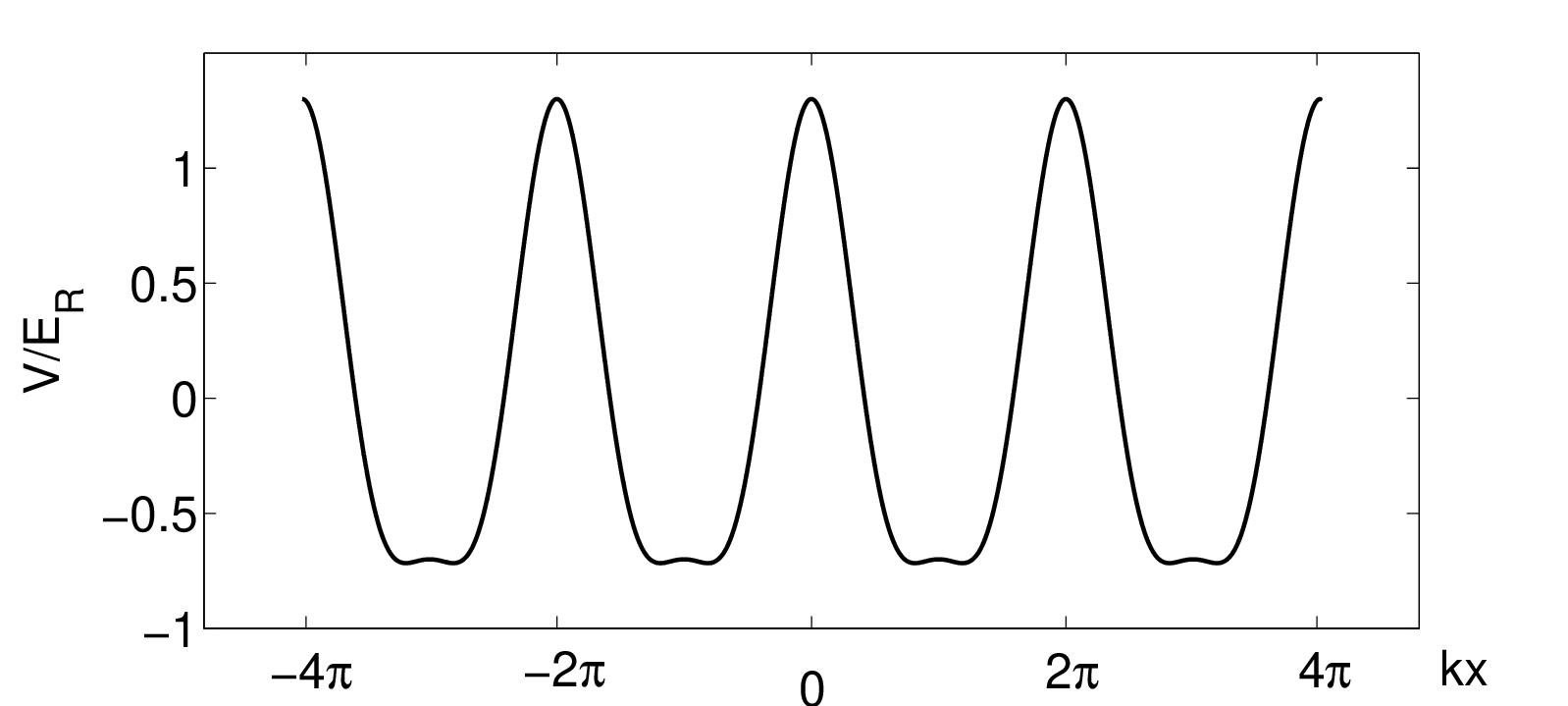}
\label{fig:fig1b}}
\caption{
\label{fig:fig1}
(Color online) (a) Optical transitions between the ground ($g$) and excited ( $e$) energy levels. Two-fold nuclear spin degeneracy
yields the manifold of states
$|g\!\uparrow\rangle$, $|g\!\downarrow\rangle$, $|e\!\uparrow\rangle$ and $|e\!\downarrow\rangle$ used to define
the potential matrix in Eq.~(\ref{eq:matrixV})
(b) The schematic view of the cross section of the bichromatic optical lattice along the
$\vec{k}_1$ direction, for the case $V'=0.2V$.(All quantities plotted are dimensionless)}
\end{figure}

All the $e-g$ transitions, shown in Fig.~\ref{fig:fig1a}, have the same frequency
$\omega_{0}=(E_{e}-E_{g})/\hbar$. We take the electric field of the lasers interacting with the atom to be
 $\vec{E}=\vec{\epsilon}e^{-i\omega t}+\vec{\epsilon}^{*}e^{i\omega t}$.
The potential in the rotating wave approximation then can be expressed as \cite{1},
\begin{equation}\label{eq:matrixV}
\hat{V}=\left( \begin{array}{cc}
 (\frac{\hbar}{2}\Delta+V_{\text{am}})\mathds{1}_{2}& -i\mathbf{\sigma}.\mathbf{\epsilon}d_{r}  \\
i\mathbf{\sigma}.\mathbf{\epsilon}d_{r} & -(\frac{\hbar}{2}\Delta+V_{\text{am}})\mathds{1}_{2} \end{array} \right),
 \end{equation}
where  $\Delta= \omega-\omega_{0}$ and $d_{r}$  is the reduced dipole moment of the atom \cite{landau}.

For the two-dimensional triangular lattice, following forms are assumed for the elements of the potential matrix,
\begin{eqnarray}
  d_{r} \mathbf{\epsilon}&=&[V\delta,V\cos(\vec{r}.\vec{k}_{1}), V\cos(\vec{r}.\vec{k}_{2})], \label{eq:dipoleElement}\\
\frac{\hbar}{2}\Delta+V_{\text{am}}(\vec{r})&=&V\cos(\vec{r}.(\vec{k}_{1}+\vec{k}_{2})) \label{eq:thetaTerm}
\end{eqnarray}
with $\vec{k}_{1}=k(1,0,0)$ and $\vec{k}_{2}=k(\cos(\theta),\sin(\theta),0)$.
Since $\omega\simeq\omega_{0}$, we have $k\simeq2 \pi / \lambda_{0}$ with
$\lambda_{0}=578\,$ nm the wavelength of the $e-g$ transition.
The spatial dependence of $V_{\text{am}}$ is set by a standing
wave at the antimagic wavelength $\lambda_{\text{am}}$
which creates a state-dependent potential with $ |\vec{k}_{1}+\vec{k}_{2}|=4 \pi / \lambda_{\text{am}}$
 that leads  to $\theta=2 \arccos(2\pi/3)$. For simplicity,
in all the following discussions we fix $\theta=2\pi/3$ and define $a\equiv4\pi/(\sqrt{3} k)$.
 Therefore the optical potential has the symmetry of a triangular
 lattice with lattice vectors $\vec{a}_{1}=(\sqrt{3}/2,-1/2)a$ and $\vec{a}_{2}=(0,1)a$.
 The Hamiltonian Eq. (\ref{hamiltonian}) is invariant
under translation by the lattice vectors $\vec{a}_{i}$. Therefore using the
associated reciprocal vectors  we can define the
Bloch Hamiltonian,
\begin{equation}\label{bloch}
  \hat{H}_{\vec{k}} = \frac{(\hat{p}+\hbar \vec{k})^{2}}{2 m}\mathds{1}_{N}+V^{\prime}(\vec{r}),
\end{equation}
 here $\vec{k}$ is the conserved momentum. $V^{\prime} = U^{-1} V U$
 where $U = 2^{-1/2}(\mathds{1}_{4}-i\Sigma_{3}\sigma_{2})$ is
 a unitary transformation. The potential $V^{\prime}$ takes the following form:
 \begin{equation}
   V^{\prime} = c_{1}\Sigma_{1}+ c_{2} \Sigma_{2}\sigma_{3} + c_{12} \Sigma_{3} +\delta \Sigma_{2}\sigma_{1}.
 \end{equation}

  In the absence of
the bichromatic lattice the Hamiltonian in Eq.~(\ref{bloch}) is invariant
under the half translational symmetries $T_{1} = \Sigma_{2}T_{\vec{a}_{1}/2}$ and
$T_{2} = \Sigma_{1}\sigma_{3}T_{\vec{a}_{2}/2}$ where $T_{\vec{a}_{i}}$ is translation operator
along the lattice vector $\vec{a}_{i}$.

 In order to examine the robustness of the $\mathbb{Z}_{2}$
topological insulator to strong perturbations we introduce bichromatic deformations. For simplicity, we do not change the angular
orientation of the lattice so that the angular variable
$\theta$ is not affected by the deformation. This can be accomplished by not
deforming the element introduced in Eq.~(\ref{eq:thetaTerm}). In addition,
we do not deform the first component in the element Eq.~(\ref{eq:dipoleElement}).
The $\delta$ dependent component plays the
role of the spin orbit coupling to hybridize the lower bands of the two optical flux lattices. We
choose not to perturb this essential process that mixes these bands with
the non-trivial Chern numbers.

The bichromatic deformation is introduced in the remaining two terms
such that
 \begin{eqnarray*}\label{potential}
 d_{r}\vec{\epsilon}&=&[V\delta,V\cos(\vec{r}.\vec{k}_{1})+V^{\prime}\cos(2\vec{r}.\vec{k}_{1}+\phi),\\
 & &V\cos(\vec{r}.\vec{k}_{2})+V^{\prime}\cos(2\vec{r}.\vec{k}_{2}+\phi)],\label{}\\
 \frac{\hbar}{2}\Delta+V_{\text{am}}(\vec{r})&=&V\cos(\vec{r}.(\vec{k}_{1}+\vec{k}_{2}))\\
 & &+V^{\prime\prime}\cos(2\vec{r}.(\vec{k}_{1}+\vec{k}_{2})+\phi').
\end{eqnarray*}
Here we examine the  case where $\phi=\phi'=0$. This model is
 unitarily equivalent to the superposition of two bichromatic optical flux lattices that is coupled
 with a spin-orbit type interaction \cite{z2cooper}. Such
 a configuration is illustrated in the Fig.~\ref{fig:fig1b} which is the cross section
 view of the potential along the $\vec{k}_1$ direction. We
explore its effect on the topological character of the insulator in the next section.
\section{Results} \label{sec:results}
\begin{figure}[t]
\subfigure[ ]{
\includegraphics[width=0.4\textwidth]{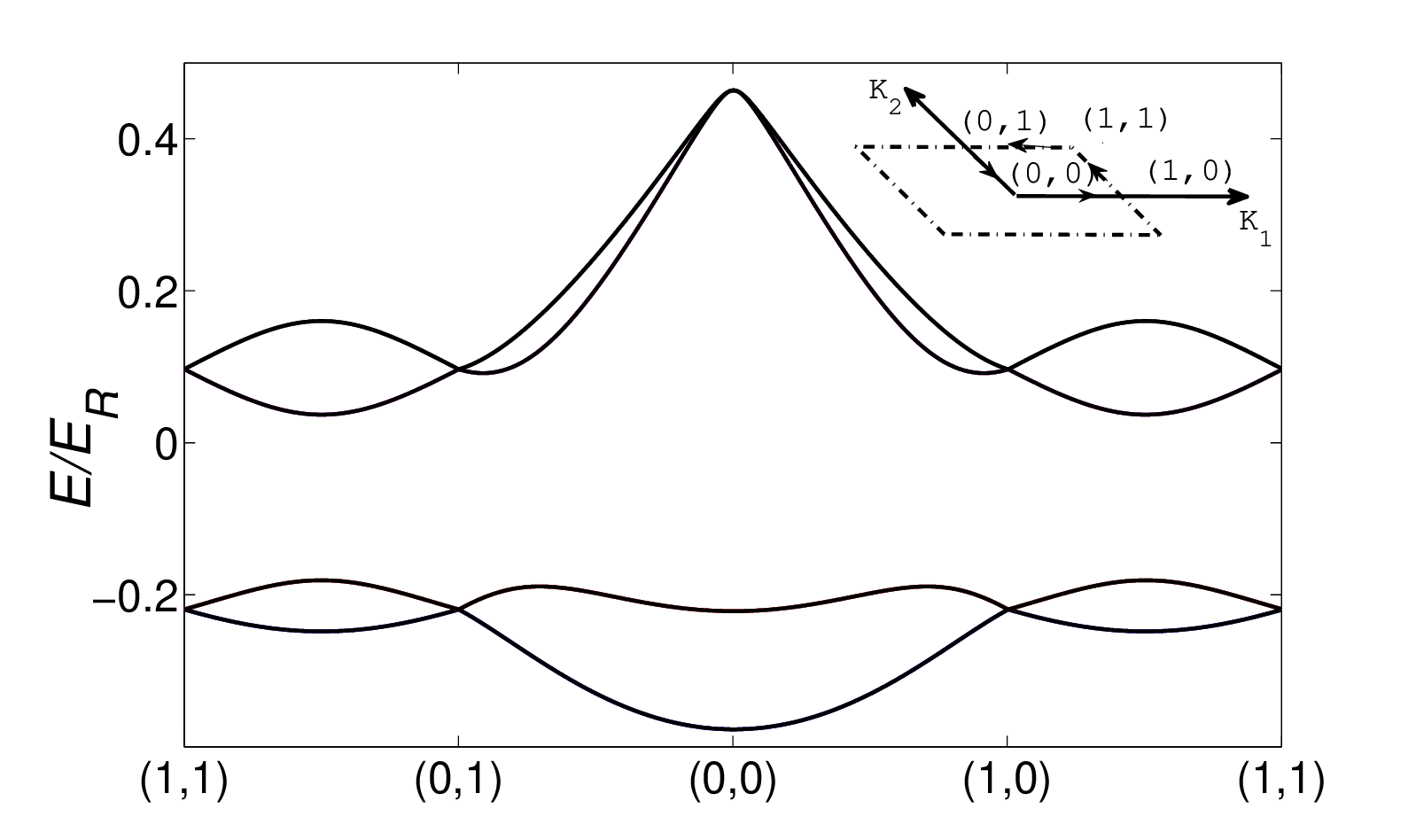}
\label{fig:fig2a}}
\qquad
\subfigure[ ]{
\includegraphics[width=0.4\textwidth]{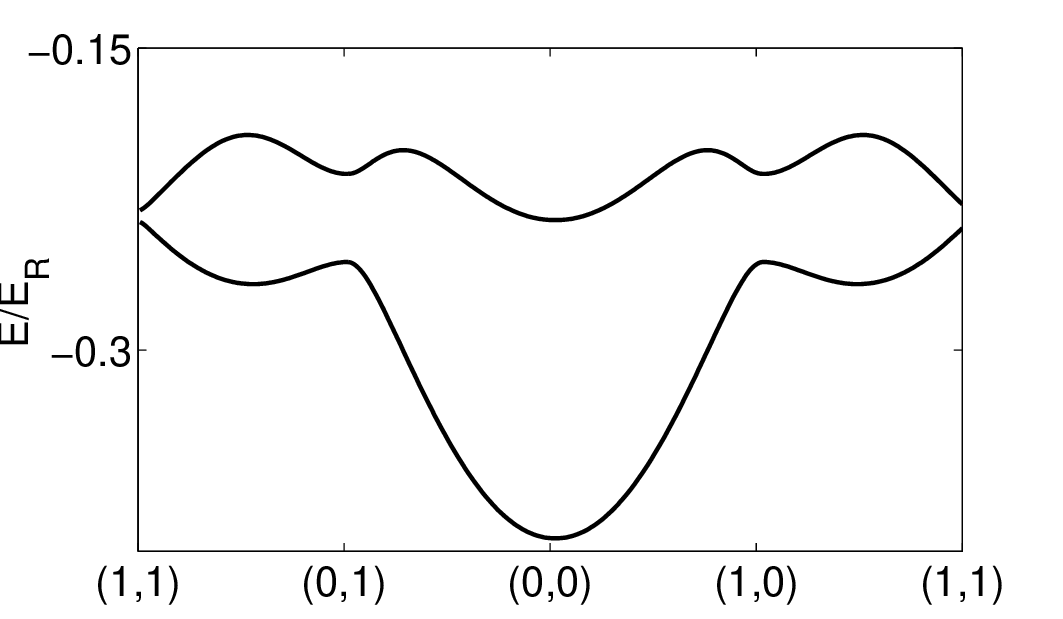}
\label{fig:fig2b}}
\qquad
\subfigure[ ]{
\includegraphics[width=0.4\textwidth]{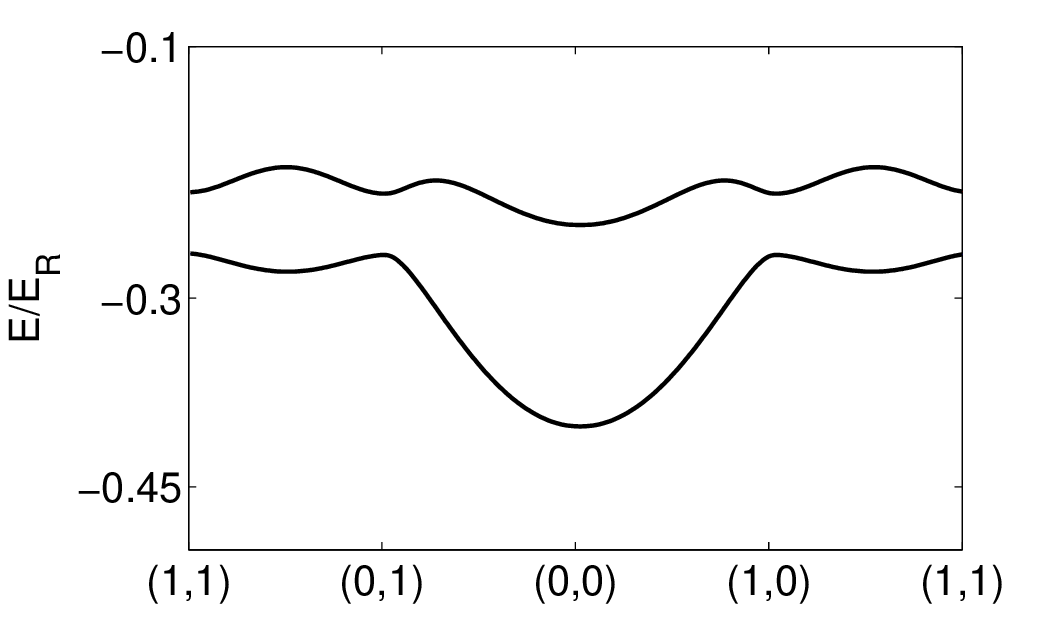}
\label{fig:fig2c}}
\caption{
\label{fig:fig2}
(Color online)
  Lowest energy bands for the lattice with $V=0.5 E_{R}$  are shown for the cases  of
(a) $V^\prime = V^{\prime \prime} =0$
and (b) $V^\prime=0.5V$ and $V^{\prime \prime} = 0$ (c) $V^\prime=0.5 V $and $V^{\prime\prime}=0.5 V$.
The \vec{k}-points are labeled as $(\vec{\Gamma}_{pq}=(p \vec{k}_{1}+q \vec{k}_{2})/2)$
 and indicated in the figure by $(p,q)$ with $p,q\in\{0,1\}$.
 At the special points (1,0), (1,1) and (0,1) the fourfold degeneracy
 is reduced to twofold by the presence of bichromatic lattice potential. }
\end{figure}

We determine the topological character of the insulator directly using the band structure
to calculate the $\mathds{Z}_{2}$ invariant.
The Hamiltonian in Eq.~(\ref{hamiltonian}) can be numerically diagonalized using plane-wave method.
The band structure is shown in Fig.~\ref{fig:fig2}. The Lowest twofold degenerate energy bands for the lattice without
disorder Fig.~\ref{fig:fig2a} has no gap between first and second bands. As we introduce the bichromatic lattice
along different direction the fourfold degeneracies at special points in k-space
are lifted.  For the cases  when the potential strength are as large as
 $V^\prime=0.5V$ and $V^{\prime\prime}=0.5 V$, Fig.~\ref{fig:fig2b} shows a gap opening
 between first and second bands but it is very narrow.  Fig.~\ref{fig:fig2b} shows
 a gap opening for the case $V^\prime=0.5 V $and $V^{\prime\prime}=0.5 V$.

In order to investigate the topological state of the system
we use the method in Ref. \cite{kane}. There are other methods
to determine the topological characteristic of the systems where
the detailed discussion and the relevant mathematical methods and
the advantages and shortcoming of different approaches on the calculation of
$\mathds{Z}_{2}$ invariant can be found in the literature \cite{qi,fukui,kane, Roy, Ryu,Yu}.
In the case of inversion symmetric materials, finding the $\mathds{Z}_{2}$
can be done using the Fu and Kane method \cite{fu}. Since the  parity and time-reversal operators
commute, we only need to check the parity eigenvalues in four time-reversal invariant points in $k$-space for all
bands below the gap. The result for the system, without bichromatic deformation, turns out to be non-trivial
so that it is either a $\mathds{Z}_{2}$ insulator or a quantum spin Hall insulator, depending on the presence
or absence of $\delta$ coupling, respectively \cite{z2cooper}.
\begin{figure}[t]
\subfigure[ ]{
\includegraphics[width=0.4\textwidth]{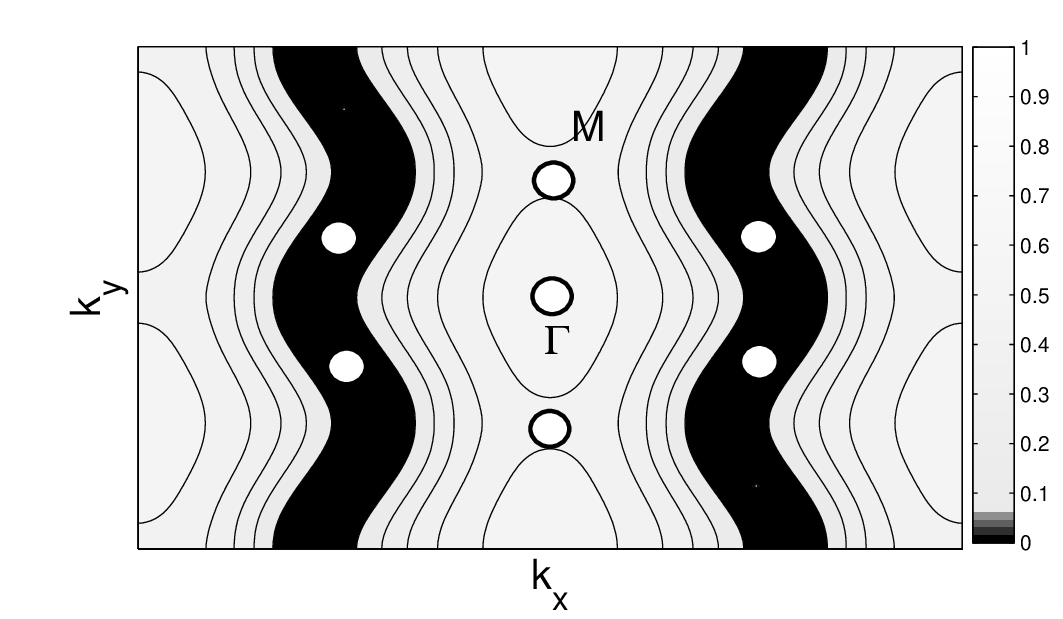}
\label{fig:fig3a}}
\subfigure[ ]{
\includegraphics[width=0.4\textwidth]{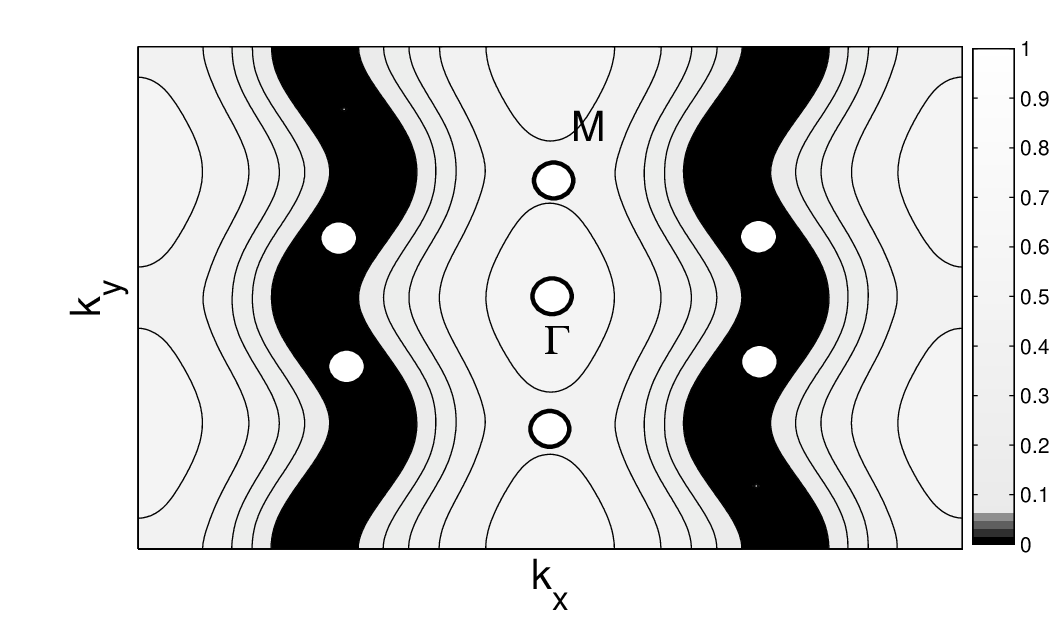}
\label{fig:fig3b}}
\subfigure[ ]{
\includegraphics[width=0.4\textwidth]{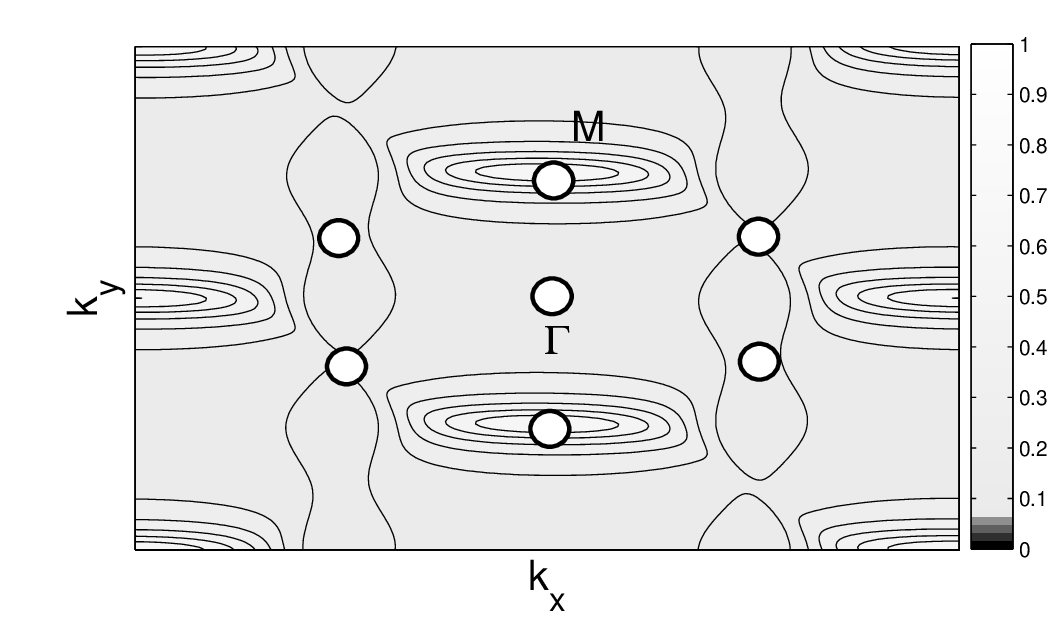}
\label{fig:fig3c}}
\caption{
\label{fig:fig3}
(Color online)
 The pfaffian $|Pf(\vec{k})|$ in \vec{k}-space for different potentials in Eq.~(\ref{potential}).
 $\Gamma$ and $M$ are high symmetry points.
 (a)  $V'=0$ and $V''=0$ are used to calculate the $|Pf(\vec{k})|$ of the  the lowest twofold degenerate two bands. (b) $|Pf(\vec{k})|$
 for $V^\prime=0.5V$ and $V^{\prime\prime}=0$. As one goes along the path $C$ which includes the half of the first Brillouin zone
 one cut the line of zeros of $|Pf(\vec{k})|$ twice.
 This shows that system is still in non trivial topological state, which is the case for Fig. \ref{fig:fig3a} as well. (c)
 $|Pf(\vec{k})|$ for lowest 2 bands for $V^\prime=0.5V$ and $V^{\prime\prime}=0.5V$ where
 there is no  line of zeros therefore the system has trivial topology. }
\end{figure}


The Hilbert space of the time reversal Hamiltonian
can be divided into two subspaces which can be called
as even and odd  depending on the relation between
the vectors and their time reversal partners.
In the even subspace the Hamiltonian
transforms under the time reversal operator as
$H_{k} = \Theta H_{k}\Theta^{-1}$ and therefore the
eigenvectors have this property that the  $\Theta |v_{i}>$
is equivalent to $|v_{i}>$ up to a $U(2)$ rotation, here  $i$ is the band index. The four
special points in half of the 2-D Brillouin zone
which are time reversal invariant points belong to even subspace.

The eigenfunctions $ |v_{i}>$  are orthogonal to their time reversal partner $\Theta |v_{i}>$ in
odd subspace.
The zeros of the Pfaffian $(Pf(\vec{k}))$ for the overlap matrix elements $\omega_{ij} = < v_{i}|\Theta|v_{j}> $ of the
time reversal operator is related to the
topological characteristic of the Hamiltonian. $|Pf(\vec{k})|$ vanishes in the odd space whereas
in the even subspace
is equal to 1 up to a $U(2)$ rotation.
The Pfaffian of a skew symmetric  matrix $\omega_{2 \times 2}$
is its $\omega_{12}$ element and it has more complicated
form for higher number of bands.  Therefore the task of finding the topological characteristic
of the Hamiltonain reduces to finding the complex zeros
of the Pfaffian. The zeros can occur either in pairs of points in the Brillouin zone
or on the lines depending on the symmetry of the system. Odd number of pairs of complex zeros
or odd number of two times cutting the line of zeros as we go through half the Brillouin zone,
indicates that we have non trivial topological insulator and
even is the sign of trivial insulator \cite{kane}.

In order to find the zeros of Pfaffian for
the bichromatic lattice we plot directly the
absolute value of the Pfaffian. Since the numerical diagonalization of the Hamiltonian
do not return in general the wave functions with the time reversal invariance,
 we impose this constraint  in the special points of the Brillouin zone
using the following relation:
\begin{equation}
|v_{i}(k)> = \Theta |v_{i-1}(k)>.
\end{equation}

The Pfaffian for lowest bands of the band structure in  Fig.~\ref{fig:fig2}
are shown in  Fig.~\ref{fig:fig3}. In the absence of the
bichromatic lattice Fig.~\ref{fig:fig2a} the zeros of Pfaffian $|Pf(k)|$ occur along a line Fig.~\ref{fig:fig3a}.
For the case when $V^{\prime} = 0$ and $V^{\prime\prime} = 0.5 V$
the system holds its nontrivial topology Fig. \ref{fig:fig3b} when we
consider two lowest twofold degenerate bands Fig.~\ref{fig:fig2b} of Hamiltonian Eq. (\ref{hamiltonian}).
 On the other hand for the case  $V^{\prime} = 0.5V$ and $V^{\prime\prime} = 0.5 V$ since we break the half translational symmetry of the Hamiltonian Eq. (\ref{hamiltonian}) Ref.~\cite{z2cooper},
 there is a gap opening
between the first and second band Fig. \ref{fig:fig2c} which  allows us to
discuss the topological characteristic of the lowest band as well.
The line of zero  vanishes for the lowest band, which means we have trivial insulator Fig.~\ref{fig:fig3c}.
However the lowest two bands together still keep their non-trivial characteristic and we saw similar pattern
as in Fig.~\ref{fig:fig3b} for this case.
The lowest band for the case $V^{\prime} = 0.5 V$ and $V^{\prime\prime} = 0 $ is also trivial but the gap
between the lowest bands is very narrow Fig.~\ref{fig:fig2b}.

\section{Conclusion} \label{sec:conclusion}
We considered the $\mathds{Z}_{2}$ topological insulator of ultracold atomic system on a triangular
optical lattice, introduced in Ref.~\cite{z2cooper}, deformed globally by bichromatic potentials.
We examined the effect of this deformation which can be as large as the insulating gap, on
the topological character of the system.
We found that the system retains its topological character robustly against this kind of
 perturbation  for two lowest twofold degenerate bands. However
 the lowest twofold degenerate band has trivial topology.
   The large insulating gap of such topological bichromatic insulators with deep
 bichromatic deformation can be used for disorder and relativistic dynamics studies in the nearly
 free electron regime.

\section*{Acknowledgments}
A.~K.~A and \"{O}.~E.~M. acknowledge financial support from T\"UB\.{I}TAK Project No.  112T974.


\begin{thebibliography}{30}
\expandafter\ifx\csname natexlab\endcsname\relax\def\natexlab#1{#1}\fi
\expandafter\ifx\csname bibnamefont\endcsname\relax
  \def\bibnamefont#1{#1}\fi
\expandafter\ifx\csname bibfnamefont\endcsname\relax
  \def\bibfnamefont#1{#1}\fi
\expandafter\ifx\csname citenamefont\endcsname\relax
  \def\citenamefont#1{#1}\fi
\expandafter\ifx\csname url\endcsname\relax
  \def\url#1{\texttt{#1}}\fi
\expandafter\ifx\csname urlprefix\endcsname\relax\def\urlprefix{URL }\fi
\providecommand{\bibinfo}[2]{#2}
\providecommand{\eprint}[2][]{\url{#2}}

\bibitem[{\citenamefont{Hasan and Kane}(2010)}]{hasan}
\bibinfo{author}{\bibfnamefont{M.~Z.} \bibnamefont{Hasan}} \bibnamefont{and}
  \bibinfo{author}{\bibfnamefont{C.~L.} \bibnamefont{Kane}},
  \bibinfo{journal}{Rev. Mod. Phys.} \textbf{\bibinfo{volume}{82}},
  \bibinfo{pages}{3045} (\bibinfo{year}{2010}).

\bibitem[{\citenamefont{Zhang}(2011)}]{qi}
\bibinfo{author}{\bibfnamefont{X.~L.~Q. S.~C.} \bibnamefont{Zhang}},
  \bibinfo{journal}{Rev. Mod. Phys.} \textbf{\bibinfo{volume}{83}},
  \bibinfo{pages}{1057} (\bibinfo{year}{2011}).

\bibitem[{\citenamefont{Moore}(2010)}]{moore2010}
\bibinfo{author}{\bibfnamefont{J.~E.} \bibnamefont{Moore}},
  \bibinfo{journal}{Nature} \textbf{\bibinfo{volume}{464}},
  \bibinfo{pages}{194} (\bibinfo{year}{2010}).

\bibitem[{\citenamefont{Nayak et~al.}(2008)\citenamefont{Nayak, Simon, Stern,
  Freedman, , and Sarma}}]{nayak}
\bibinfo{author}{\bibfnamefont{C.}~\bibnamefont{Nayak}},
  \bibinfo{author}{\bibfnamefont{S.~H.} \bibnamefont{Simon}},
  \bibinfo{author}{\bibfnamefont{A.}~\bibnamefont{Stern}},
  \bibinfo{author}{\bibfnamefont{M.}~\bibnamefont{Freedman}}, ,
  \bibnamefont{and} \bibinfo{author}{\bibfnamefont{S.~D.} \bibnamefont{Sarma}},
  \bibinfo{journal}{Rev. Mod. Phys.} \textbf{\bibinfo{volume}{80}},
  \bibinfo{pages}{1083} (\bibinfo{year}{2008}).

\bibitem[{\citenamefont{Hu et~al.}(2011)\citenamefont{Hu, Kargarian, and
  Fiete}}]{lattices1}
\bibinfo{author}{\bibfnamefont{X.}~\bibnamefont{Hu}},
  \bibinfo{author}{\bibfnamefont{M.}~\bibnamefont{Kargarian}},
  \bibnamefont{and} \bibinfo{author}{\bibfnamefont{G.~A.} \bibnamefont{Fiete}},
  \bibinfo{journal}{Phys. Rev. B} \textbf{\bibinfo{volume}{84}},
  \bibinfo{pages}{155116} (\bibinfo{year}{2011}).

\bibitem[{\citenamefont{Weeks and Franz}(2010)}]{lattices2}
\bibinfo{author}{\bibfnamefont{C.}~\bibnamefont{Weeks}} \bibnamefont{and}
  \bibinfo{author}{\bibfnamefont{M.}~\bibnamefont{Franz}},
  \bibinfo{journal}{Phys. Rev. B} \textbf{\bibinfo{volume}{82}},
  \bibinfo{pages}{085310} (\bibinfo{year}{2010}).

\bibitem[{\citenamefont{Guo and Franz}(2009{\natexlab{a}})}]{lattices3}
\bibinfo{author}{\bibfnamefont{H.-M.} \bibnamefont{Guo}} \bibnamefont{and}
  \bibinfo{author}{\bibfnamefont{M.}~\bibnamefont{Franz}},
  \bibinfo{journal}{Phys. Rev. B} \textbf{\bibinfo{volume}{80}},
  \bibinfo{pages}{113102} (\bibinfo{year}{2009}{\natexlab{a}}).

\bibitem[{\citenamefont{Guo and Franz}(2009{\natexlab{b}})}]{lattices4}
\bibinfo{author}{\bibfnamefont{H.-M.} \bibnamefont{Guo}} \bibnamefont{and}
  \bibinfo{author}{\bibfnamefont{M.}~\bibnamefont{Franz}},
  \bibinfo{journal}{Phys. Rev. Lett.} \textbf{\bibinfo{volume}{103}},
  \bibinfo{pages}{20680} (\bibinfo{year}{2009}{\natexlab{b}}).

\bibitem[{\citenamefont{L.~Fu and Mele}(2007)}]{lattices5}
\bibinfo{author}{\bibfnamefont{C.~L.~K.} \bibnamefont{L.~Fu}} \bibnamefont{and}
  \bibinfo{author}{\bibfnamefont{E.~J.} \bibnamefont{Mele}},
  \bibinfo{journal}{Phys. Rev. Lett.} \textbf{\bibinfo{volume}{98}},
  \bibinfo{pages}{106803} (\bibinfo{year}{2007}).

\bibitem[{\citenamefont{Schubert et~al.}(2012)\citenamefont{Schubert, Fehske,
  Fritz, and Vojta}}]{schubert}
\bibinfo{author}{\bibfnamefont{G.}~\bibnamefont{Schubert}},
  \bibinfo{author}{\bibfnamefont{H.}~\bibnamefont{Fehske}},
  \bibinfo{author}{\bibfnamefont{L.}~\bibnamefont{Fritz}}, \bibnamefont{and}
  \bibinfo{author}{\bibfnamefont{M.}~\bibnamefont{Vojta}},
  \bibinfo{journal}{Phys. Rev. B.} \textbf{\bibinfo{volume}{85}},
  \bibinfo{pages}{201105} (\bibinfo{year}{2012}).

\bibitem[{\citenamefont{B\'eri and Cooper}(2011)}]{z2cooper}
\bibinfo{author}{\bibfnamefont{B.}~\bibnamefont{B\'eri}} \bibnamefont{and}
  \bibinfo{author}{\bibfnamefont{N.~R.} \bibnamefont{Cooper}},
  \bibinfo{journal}{Phys. Rev. Lett.} \textbf{\bibinfo{volume}{107}},
  \bibinfo{pages}{145301} (\bibinfo{year}{2011}).

\bibitem[{\citenamefont{Juzelians et~al.}(2010)\citenamefont{Juzelians,
  Ruseckas, and Dalibard}}]{imprint1}
\bibinfo{author}{\bibfnamefont{G.}~\bibnamefont{Juzelians}},
  \bibinfo{author}{\bibfnamefont{J.}~\bibnamefont{Ruseckas}}, \bibnamefont{and}
  \bibinfo{author}{\bibfnamefont{J.}~\bibnamefont{Dalibard}},
  \bibinfo{journal}{Phys. Rev. A} \textbf{\bibinfo{volume}{81}},
  \bibinfo{pages}{053403} (\bibinfo{year}{2010}).

\bibitem[{\citenamefont{Goldman et~al.}(2010)\citenamefont{Goldman, Satija,
  Nikolic, Bermudez, Martin-Delgado, Lewenstein, and Spielman}}]{imprint2}
\bibinfo{author}{\bibfnamefont{N.}~\bibnamefont{Goldman}},
  \bibinfo{author}{\bibfnamefont{I.}~\bibnamefont{Satija}},
  \bibinfo{author}{\bibfnamefont{P.}~\bibnamefont{Nikolic}},
  \bibinfo{author}{\bibfnamefont{A.}~\bibnamefont{Bermudez}},
  \bibinfo{author}{\bibfnamefont{M.~A.} \bibnamefont{Martin-Delgado}},
  \bibinfo{author}{\bibfnamefont{M.}~\bibnamefont{Lewenstein}},
  \bibnamefont{and} \bibinfo{author}{\bibfnamefont{I.~B.}
  \bibnamefont{Spielman}}, \bibinfo{journal}{Phys. Rev. Lett.}
  \textbf{\bibinfo{volume}{105}}, \bibinfo{pages}{255302}
  (\bibinfo{year}{2010}).

\bibitem[{\citenamefont{Lin et~al.}(2011)\citenamefont{Lin, Jim\'enez-Garc\'ia,
  and Spielman}}]{spielman}
\bibinfo{author}{\bibfnamefont{Y.-J.} \bibnamefont{Lin}},
  \bibinfo{author}{\bibfnamefont{K.}~\bibnamefont{Jim\'enez-Garc\'ia}},
  \bibnamefont{and} \bibinfo{author}{\bibfnamefont{I.~B.}
  \bibnamefont{Spielman}}, \bibinfo{journal}{Nature}
  \textbf{\bibinfo{volume}{471}}, \bibinfo{pages}{83} (\bibinfo{year}{2011}).

\bibitem[{\citenamefont{Witthaut et~al.}(2011)\citenamefont{Witthaut, Salger,
  Kling, Grossert, and Weitz}}]{withaut}
\bibinfo{author}{\bibfnamefont{D.}~\bibnamefont{Witthaut}},
  \bibinfo{author}{\bibfnamefont{T.}~\bibnamefont{Salger}},
  \bibinfo{author}{\bibfnamefont{S.}~\bibnamefont{Kling}},
  \bibinfo{author}{\bibfnamefont{C.}~\bibnamefont{Grossert}}, \bibnamefont{and}
  \bibinfo{author}{\bibfnamefont{M.}~\bibnamefont{Weitz}},
  \bibinfo{journal}{Phys. Rev. A} \textbf{\bibinfo{volume}{84}},
  \bibinfo{pages}{033601} (\bibinfo{year}{2011}).

\bibitem[{\citenamefont{Mazza et~al.}(2012)\citenamefont{Mazza, Bermudez,
  Goldman, Rizzi, Martin-Delgado, and Lewenstein}}]{mazza2012}
\bibinfo{author}{\bibfnamefont{L.}~\bibnamefont{Mazza}},
  \bibinfo{author}{\bibfnamefont{A.}~\bibnamefont{Bermudez}},
  \bibinfo{author}{\bibfnamefont{N.}~\bibnamefont{Goldman}},
  \bibinfo{author}{\bibfnamefont{M.}~\bibnamefont{Rizzi}},
  \bibinfo{author}{\bibfnamefont{M.~A.} \bibnamefont{Martin-Delgado}},
  \bibnamefont{and}
  \bibinfo{author}{\bibfnamefont{M.}~\bibnamefont{Lewenstein}},
  \bibinfo{journal}{New Jour. Phys.} \textbf{\bibinfo{volume}{14}},
  \bibinfo{pages}{015007} (\bibinfo{year}{2012}).

\bibitem[{\citenamefont{Aubry and Andr\'e}(1980)}]{aubry}
\bibinfo{author}{\bibfnamefont{S.}~\bibnamefont{Aubry}} \bibnamefont{and}
  \bibinfo{author}{\bibfnamefont{G.}~\bibnamefont{Andr\'e}},
  \bibinfo{journal}{Ann. Israel Phys. Soc.} \textbf{\bibinfo{volume}{3}},
  \bibinfo{pages}{133} (\bibinfo{year}{1980}).

\bibitem[{\citenamefont{Roati et~al.}(2008)\citenamefont{Roati, D'Errico,
  Fallani, Fattori, Fort, Zaccanti, Modugno, and Inguscio}}]{roati}
\bibinfo{author}{\bibfnamefont{G.}~\bibnamefont{Roati}},
  \bibinfo{author}{\bibfnamefont{C.}~\bibnamefont{D'Errico}},
  \bibinfo{author}{\bibfnamefont{L.}~\bibnamefont{Fallani}},
  \bibinfo{author}{\bibfnamefont{M.}~\bibnamefont{Fattori}},
  \bibinfo{author}{\bibfnamefont{C.}~\bibnamefont{Fort}},
  \bibinfo{author}{\bibfnamefont{M.}~\bibnamefont{Zaccanti}},
  \bibinfo{author}{\bibfnamefont{G.~M.~M.} \bibnamefont{Modugno}},
  \bibnamefont{and} \bibinfo{author}{\bibfnamefont{M.}~\bibnamefont{Inguscio}},
  \bibinfo{journal}{Nature} \textbf{\bibinfo{volume}{453}},
  \bibinfo{pages}{895898} (\bibinfo{year}{2008}).

\bibitem[{\citenamefont{Deissler et~al.}(2010)\citenamefont{Deissler, Zaccanti,
  Roati, D’Errico, Fattori, Modugno, Modugno, , and Inguscio}}]{deissler2010}
\bibinfo{author}{\bibfnamefont{B.}~\bibnamefont{Deissler}},
  \bibinfo{author}{\bibfnamefont{M.}~\bibnamefont{Zaccanti}},
  \bibinfo{author}{\bibfnamefont{G.}~\bibnamefont{Roati}},
  \bibinfo{author}{\bibfnamefont{C.}~\bibnamefont{D’Errico}},
  \bibinfo{author}{\bibfnamefont{M.}~\bibnamefont{Fattori}},
  \bibinfo{author}{\bibfnamefont{M.}~\bibnamefont{Modugno}},
  \bibinfo{author}{\bibfnamefont{G.}~\bibnamefont{Modugno}}, ,
  \bibnamefont{and} \bibinfo{author}{\bibfnamefont{M.}~\bibnamefont{Inguscio}},
  \bibinfo{journal}{Nature Phys.} \textbf{\bibinfo{volume}{6}},
  \bibinfo{pages}{354} (\bibinfo{year}{2010}).

\bibitem[{\citenamefont{Mei et~al.}(2012)\citenamefont{Mei, Zhu, Zhang, Oh, and
  Goldman}}]{mei2012}
\bibinfo{author}{\bibfnamefont{F.}~\bibnamefont{Mei}},
  \bibinfo{author}{\bibfnamefont{S.-L.} \bibnamefont{Zhu}},
  \bibinfo{author}{\bibfnamefont{A.-M.} \bibnamefont{Zhang}},
  \bibinfo{author}{\bibfnamefont{C.~H.} \bibnamefont{Oh}}, \bibnamefont{and}
  \bibinfo{author}{\bibfnamefont{N.}~\bibnamefont{Goldman}},
  \bibinfo{journal}{Phys. Rev. A} \textbf{\bibinfo{volume}{85}},
  \bibinfo{pages}{013638} (\bibinfo{year}{2012}).

\bibitem[{\citenamefont{Lang et~al.}(2012)\citenamefont{Lang, Cai, and
  Chen}}]{lang2012}
\bibinfo{author}{\bibfnamefont{L.-J.} \bibnamefont{Lang}},
  \bibinfo{author}{\bibfnamefont{X.}~\bibnamefont{Cai}}, \bibnamefont{and}
  \bibinfo{author}{\bibfnamefont{S.}~\bibnamefont{Chen}},
  \bibinfo{journal}{Phys. Rev. Lett.} \textbf{\bibinfo{volume}{108}},
  \bibinfo{pages}{220401} (\bibinfo{year}{2012}).

\bibitem[{\citenamefont{Kane and Mele}(2005)}]{kane}
\bibinfo{author}{\bibfnamefont{C.}~\bibnamefont{Kane}} \bibnamefont{and}
  \bibinfo{author}{\bibfnamefont{E.}~\bibnamefont{Mele}},
  \bibinfo{journal}{Phys. Rev. lett.} \textbf{\bibinfo{volume}{95}},
  \bibinfo{pages}{146802} (\bibinfo{year}{2005}).

\bibitem[{\citenamefont{Gerbier and Dalibard}(2010)}]{dalibard}
\bibinfo{author}{\bibfnamefont{F.}~\bibnamefont{Gerbier}} \bibnamefont{and}
  \bibinfo{author}{\bibfnamefont{J.}~\bibnamefont{Dalibard}},
  \bibinfo{journal}{New Jour.~Phys.} \textbf{\bibinfo{volume}{12}},
  \bibinfo{pages}{033007} (\bibinfo{year}{2010}).

\bibitem[{\citenamefont{Cohen-Tannoudji
  et~al.}(1992)\citenamefont{Cohen-Tannoudji, Dupont-Roc, and Grynberg}}]{1}
\bibinfo{author}{\bibfnamefont{C.}~\bibnamefont{Cohen-Tannoudji}},
  \bibinfo{author}{\bibfnamefont{J.}~\bibnamefont{Dupont-Roc}},
  \bibnamefont{and} \bibinfo{author}{\bibfnamefont{G.}~\bibnamefont{Grynberg}},
  \emph{\bibinfo{title}{Atom-Photon Interactions}} (\bibinfo{publisher}{Wiley},
  \bibinfo{address}{New York}, \bibinfo{year}{1992}).

\bibitem[{\citenamefont{Landau and Lifshitz}(1958)}]{landau}
\bibinfo{author}{\bibfnamefont{L.~D.} \bibnamefont{Landau}} \bibnamefont{and}
  \bibinfo{author}{\bibfnamefont{E.~M.} \bibnamefont{Lifshitz}},
  \emph{\bibinfo{title}{Quantum Mechanics}} (\bibinfo{publisher}{Pergamon},
  \bibinfo{address}{London}, \bibinfo{year}{1958}).

\bibitem[{\citenamefont{Fukui and Hatsugai}(2007)}]{fukui}
\bibinfo{author}{\bibfnamefont{T.}~\bibnamefont{Fukui}} \bibnamefont{and}
  \bibinfo{author}{\bibfnamefont{Y.}~\bibnamefont{Hatsugai}},
  \bibinfo{journal}{J.~Phys.~Soc.~Jpn.~} \textbf{\bibinfo{volume}{76}},
  \bibinfo{pages}{053702} (\bibinfo{year}{2007}).

\bibitem[{\citenamefont{Roy}(2009)}]{Roy}
\bibinfo{author}{\bibfnamefont{R.}~\bibnamefont{Roy}}, \bibinfo{journal}{Phys.
  Rev. B} \textbf{\bibinfo{volume}{79}}, \bibinfo{pages}{195321}
  (\bibinfo{year}{2009}).

\bibitem[{\citenamefont{Ryu et~al.}(2007)\citenamefont{Ryu, Mudry, Obuse, and
  Furusaki}}]{Ryu}
\bibinfo{author}{\bibfnamefont{S.}~\bibnamefont{Ryu}},
  \bibinfo{author}{\bibfnamefont{C.}~\bibnamefont{Mudry}},
  \bibinfo{author}{\bibfnamefont{H.}~\bibnamefont{Obuse}}, \bibnamefont{and}
  \bibinfo{author}{\bibfnamefont{A.}~\bibnamefont{Furusaki}},
  \bibinfo{journal}{Phys. Rev. Lett.} \textbf{\bibinfo{volume}{99}},
  \bibinfo{pages}{116601} (\bibinfo{year}{2007}).

\bibitem[{\citenamefont{Yu et~al.}(2011)\citenamefont{Yu, Qi, Bernevig, Fang,
  and Dai}}]{Yu}
\bibinfo{author}{\bibfnamefont{R.}~\bibnamefont{Yu}},
  \bibinfo{author}{\bibfnamefont{X.}~\bibnamefont{Qi}},
  \bibinfo{author}{\bibfnamefont{A.}~\bibnamefont{Bernevig}},
  \bibinfo{author}{\bibfnamefont{Z.}~\bibnamefont{Fang}}, \bibnamefont{and}
  \bibinfo{author}{\bibfnamefont{X.}~\bibnamefont{Dai}},
  \bibinfo{journal}{Phys. Rev. B} \textbf{\bibinfo{volume}{84}},
  \bibinfo{pages}{075119} (\bibinfo{year}{2011}).

\bibitem[{\citenamefont{Fu and Kane}(2006)}]{fu}
\bibinfo{author}{\bibfnamefont{L.}~\bibnamefont{Fu}} \bibnamefont{and}
  \bibinfo{author}{\bibfnamefont{C.~L.} \bibnamefont{Kane}},
  \bibinfo{journal}{Phys. Rev. B} \textbf{\bibinfo{volume}{74}},
  \bibinfo{pages}{195312} (\bibinfo{year}{2006}).

\end{thebibliography}

\end{document}